\documentclass[conference]{IEEEtran}
\usepackage{cite}
\usepackage{amsmath}
\usepackage{float}
\usepackage{amssymb,amsfonts}
\usepackage{algorithmic}
\usepackage{comment}
\usepackage{lipsum}
\usepackage{graphicx}
\usepackage{textcomp}
\usepackage{siunitx}
\usepackage{xcolor, soul}
\usepackage{hyperref}
\usepackage{cleveref}
\usepackage{enumitem}
\usepackage[acronym]{glossaries}
\usepackage[most]{tcolorbox}
\usepackage[english]{babel}
\tcbuselibrary{theorems}
\def\BibTeX{{\rm B\kern-.05em{\sc i\kern-.025em b}\kern-.08em
    T\kern-.1667em\lower.7ex\hbox{E}\kern-.125emX}}

\makeglossaries
\newacronym{uas}{UAS}{Uncrewed Aircraft System}
\newacronym{uav}{UAV}{Uncrewed Aircraft Vehicle}
\newacronym{uavs}{UAVs}{Uncrewed Aircraft Vehicles}
\newacronym{utm}{UTM}{UAS Traffic Management}
\newacronym{evtol}{eVTOL}{electric Vertical Takeoff and Landing}
\newacronym{agl}{AGL}{Above Ground Level}
\newacronym{aam}{AAM}{Advanced Air Mobility}
\newacronym{faa}{FAA}{Federal Aviation Administration}
\newacronym{atc}{ATC}{Air Traffic Control}
\newacronym{i28}{I28}{Innovate28}
\newacronym{uass}{UASs}{Uncrewed Aircraft Systems}
\newacronym{uid}{UID}{unique identification}
\newacronym{tcf}{TCF}{Traffic Control Framework}
\newacronym{ta}{TA}{Traffic Analytics}
\newacronym{lacc}{LACC}{Level of Autonomy in Cognitive Control}
\newacronym{lbsd}{LBSD}{Lane-Based Strategic Deconfliction}
\newacronym{ai}{AI}{Artificial Intelligence}
\newacronym{ml}{ML}{Machine Learning}
\newacronym{rl}{RL}{Reinforcement Learning}
\newacronym{tpi}{TPI}{Traffic Performance Indicator}
\newacronym{fnsd}{FNSD}{FAA-NASA Strategic Deconfliction}
\begin{document}
\title{A Traffic Control Framework for Uncrewed Aircraft Systems}

\newtheorem{assumption}{Assumption}
\newtheorem{definition}{\textbf{Definition}}

\author{\IEEEauthorblockN{Ananay Vikram Gupta}
\IEEEauthorblockA{\textit{Computer Science} \\
\textit{Georgia Institute of Technology}\\
Atlanta, United States \\
ananay@gatech.edu}
\and
\IEEEauthorblockN{Aaditya Prakash Kattekola}
\IEEEauthorblockA{\textit{Electronics \& Communication Engineering} \\
\textit{National Institute of Technology}\\
Warangal, India \\
aadityapra2910@gmail.com}
\and
\IEEEauthorblockN{Ansh Vikram Gupta}
\IEEEauthorblockA{\textit{Computer Science} \\
\textit{Rose-Hulman Institute of Technology}\\
Terre Haute, United States \\
guptaav@rose-hulman.edu}
\and
\IEEEauthorblockN{Dacharla Venkata Abhiram}
\IEEEauthorblockA{\textit{Computer Science \& Engineering} \\
\textit{IIITDM Kancheepuram}\\
Chennai, India \\
dacharlaabhiram0908@gmail.com}
\and
\IEEEauthorblockN{Kamesh Namuduri}
\IEEEauthorblockA{\textit{Electrical Engineering} \\
\textit{University of North Texas}\\
Denton, United States \\
Kamesh.Namuduri@unt.edu}
\and
\IEEEauthorblockN{Ravichandran Subramanian}
\IEEEauthorblockA{\textit{Principal Scientist} \\
\textit{Hermes Autonomous Air Mobility Solutions }\\
Delaware, United States \\
ravisub@gmail.com}
}

\maketitle

\begin{abstract}The exponential growth of \acrfull{aam} services demands assurances of safety in the airspace. This research presents a \acrfull{tcf} for developing digital flight rules for \acrfull{uas} flying in designated air corridors. The proposed \acrshort{tcf} helps model, deploy, and test \acrshort{uas} control agents, regardless of their hardware configurations.
This paper investigates the importance of digital flight rules in preventing collisions in the context of \acrshort{aam} \acrshort{tcf} is introduced as a platform for developing strategies for managing \acrshort{uas} traffic towards enhanced autonomy in the airspace. It allows for assessment and evaluation autonomous navigation, route planning, obstacle avoidance, and adaptive decision-making for \acrshort{uas}.  It also allows for the introduction and evaluation of advanced technologies including Artificial Intelligence (AI) and Machine Learning (ML) in a simulation environment before deploying them in the real world. \acrshort{tcf} can be used as a tool for comprehensive \acrshort{uas} traffic analysis, including KPI measurement. It offers flexibility for further testing and deployment, laying the foundation for improved airspace safety—a vital aspect of \acrshort{uas} technology advancement.
Finally, this paper demonstrates the capabilities of the proposed  \acrshort{tcf}in managing \acrshort{uas} traffic at intersections and its impact on overall traffic flow in air corridors, noting the bottlenecks and the inverse relationship between safety and traffic volume.

\end{abstract}


\section{Introduction}
\acrfull{aam} is a rapidly emerging area of reserach and development in the aviation industry that aims to provide effective, sustainable, and efficient means for transportation of people and goods. The demand for \acrshort{aam} services is on the rise around the world in terms of both commercial and industrial aspects however, \acrshort{aam} brings with it a pre-requisite of highest levels safety in the airspace.  \acrshort{aam} can not become a reality without the highest level of safety assurance from the aviation industry. 
In order to successfully ensure the safety of \acrshort{aam} platforms including \acrfull{uav}s, it is important to establish clear guidelines and flight rules in the airspace in which \acrfull{uas} are permitted to fly. This will prevent collisions and ensure that \acrshort{uav} do not interfere with other aircraft or people{'}s privacy. These guidelines can also allow for efficient airspace use and help reduce congestion.
Developing regulations for air control management is much harder to implement due to the lack of resources. Controlled Structural Airspaces are yet to be built, and as such researchers are not given the opportunity to analyze patterns and behaviors of \acrshort{uav} in a physical setting.
Cutting-edge technologies like autonomous navigation, \acrshort{ai}-driven route optimization, real-time obstacle avoidance, and adaptive decision-making enhance UAV operations \cite{FAA-AI2023}. \acrfull{rl} further boosts \acrshort{uav} performance. The \acrshort{ai}/\acrshort{ml} assurance framework facilitates simulation and real-world deployment of \acrshort{ai}-backed control policies.\cite{FAA-AI2023}
\vspace{-5pt}
\subsection{Contributions}
This paper presents the \acrfull{tcf} as a tool for analyzing air traffic in a controlled environment among \acrshort{uavs}. The proposed \acrshort{tcf} has also been implemented as a simulation tool. The capability of this tool in assessing  key traffic performance metrics has been demonstrated. The \acrshort{tcf} allows for modification and extension to test out more complex models or attempt to deploy the \acrshort{uav} control models used in the simulations on real drones in a controlled environment. The research and analysis can be used as a foundation for enhanced safety in the airspace, essential for the future of \acrshort{uav} technology.


\subsection{Organization of this paper}
\textcolor{green}{review}
\Cref{sec:lit rev} provides a comprehensive analysis of the recent studies and advancements within Air Traffic Management for \acrshort{uav}. \Cref{sec:airspace} is an informative analysis of the structured spaces required for Air Traffic Management. The safety and regulatory aspects of air corridors, skylanes, and air cells are presented in this section. \Cref{sec:TCF} introduces the \acrshort{tcf} of the research paper. This section introduces control fundamentals, various traffic rules such as intersection handling, and \acrlong{ta}. The results of the \acrshort{tcf} simulation are shown in \Cref{sec:sim and result} and \Cref{sec:conclusion} provides a conclusive review of the entire paper.

\section{Literature Review}
\label{sec:lit rev}
\acrshort{uas} have been in use for military purposes for over a century, but their use has spread to commercial and civilian applications in recent years. However, the increased use of \acrshort{uas} has created a need for managing the growing number of \acrshort{uas} in the sky to ensure their safe and efficient integration into airspace, similar to how \acrfull{atc} regulates crewed aircraft.
\par
This need has led to the development of \acrfull{utm} \cite{FAA-UTM2020}, which is required to guarantee the safe, secure, and efficient integration of \acrshort{uas} in the National Airspace System. \acrshort{utm} is intended for traffic management (monitoring, and enforcing traffic rules) of small \acrshort{uas} flying under 400 ft \acrfull{agl}.

\par
\acrshort{aam} platforms such as \acrfull{evtol} vehicles fly between 500 ft - 3000 ft \acrshort{agl} and are intended for transporting people and large cargo. \acrshort{aam} architecture includes \cite{NASA-Corridors2020} 
federated traffic management system also known as Provider of Services for \acrshort{aam} and Supplemental Data Service Providers. Regulations are evolving to safely and efficiently integrate \acrshort{utm} and \acrshort{aam} air traffic.
\par
The idea of air corridors is crucial in the context of \acrshort{aam}. Air corridors\cite{Muna2021corridors} are specialized paths inside the airspace that are intended for \acrshort{aam} vehicles, such as \acrshort{evtol} planes, to move between locations. These corridors are different from regular manned aircraft routes and serve as specialized paths for \acrshort{aam} vehicles.
\par
One of the major issues that air corridors address is the possibility of conflict and safety concerns brought on by the integration of \acrshort{aam} vehicles into conventional airspace systems. Potential confrontations with other airspace users, including manned and unmanned aircraft, can be reduced by designating specific air corridors. The usage of corridors can be helpful in resolving challenges of airspace availability in urban settings, which are constrained by building height, weather effects, privacy requirements, and current air traffic flows\cite{Verma2022UAM}. 
\par
Air corridors also deal with the issue of route planning optimization and ensuring effective operations for \acrshort{aam}  vehicles \cite{Muna2021corridors}. \acrshort{aam} vehicles can use optimized routes that streamline operations, cut down on travel time, and improve overall efficiency by designating specified corridors. Air corridors offer passengers and \acrshort{aam} operators a reliable and predictable framework, which enhances the appeal and feasibility of \acrshort{aam} systems.
\par
Moreover, by defining specific corridors and associated operating procedures, air corridors provide a clear framework for managing \acrshort{aam} traffic within the broader airspace ecosystem. This integration allows \acrshort{atc} and \acrshort{utm} systems to effectively monitor and regulate \acrshort{aam} operations, ensuring compliance with necessary regulations and maintaining the overall safety and reliability of the airspace system.

\par
Various researchers propose approaches for traffic control frameworks. As discussed in \cite{Verma2022UAM}, one such technique proposes a strategy to consider airspace constraints, \acrshort{uas} performance, and safety requirements, using simulation tools and advanced technologies to assess their efficacy. New policies and regulations specific to \acrshort{aam} operations are needed, and factors such as airspace structure, terrain, weather, environmental factors, and \acrshort{uas} capabilities must be considered. Safety requirements such as collision avoidance and emergency procedures must also be met.
\par
A lane-based method, introduced in \cite{sacharny2022lane}, proposes a standardized approach to coordinate \acrshort{uas} traffic and enhance predictability and safety. It utilizes clearly defined flight paths to ensure separation from other \acrshort{uas} and manned aircraft. Based on designated lanes, the \acrfull{lbsd} system enables efficient and secure \acrshort{uas} operations while maintaining safety. The study also discusses the \acrfull{fnsd} method, which uses "block rules" to regulate \acrshort{uas} traffic. However, \acrshort{lbsd} is considered more adaptable and effective. The regulatory framework for \acrshort{uas} traffic management may require further development to address the complexities and improve safety and effectiveness.
\par
The Joint Control Framework \cite{Lundberg2021JCF} describes \acrshort{utm} as a collaboration between man \& machine. It introduces \acrfull{lacc} which delineates the 6 levels of control present in human-robot interaction. A traffic simulation in \cite{Lundberg2021UTM} illustrates an advantage of \acrshort{lacc} within the context of \acrshort{utm}. This simulation utilizes 3D real-world maps and geofences. However, these papers prove a practical hypothesis. They do not specify any methodology or implementation details in managing air traffic.  
\par

In urban air freight operations within the vertiport environment,  \cite{NTRS-Vertiport2021} emphasizes the vital role of 4D flight trajectories in \acrshort{aam}. These trajectories, integrating spatial coordinates and time, enable precise route planning and obstacle avoidance, facilitated by Detect and Avoid (DAA) capabilities. DAA employs advanced sensors and algorithms to detect and track obstacles, allowing UAVs to adjust their flight paths in real time. This integration enhances coordination between aircraft and \acrshort{atc}, improving operational efficiency and safety.
\par
The \acrfull{faa} plan \cite{FAA-AAM2023} for \acrshort{aam} integration including the \acrfull{i28} program, holds significance for \acrshort{uas} traffic control. The \acrshort{faa}'s crawl-walk-run approach, coordination with \acrshort{atc}, adherence to Communication, Navigation, and Surveillance regulations, and tailored routing constructs for \acrshort{aam} resonate with \acrshort{uas} route planning. Infrastructure considerations, such as charging stations and parking zones, align with operational needs. This strategic alignment underscores how the \acrshort{aam} plan, along with the \acrshort{i28} program, informs \acrshort{uas} traffic control.
\section{Structured Airspaces}
\label{sec:airspace}
Structured airspaces such as air corridors  play a critical role in facilitating safe and efficient traversal by \acrshort{uas}s. By establishing designated routes and implementing traffic rules, structured airspaces contribute to reducing the likelihood of collisions and traffic congestion, consequently minimizing delays. These airspaces also enable \acrshort{uav}s to swiftly reach their destinations with minimal reliance on extensive environmental monitoring.\par
Further, structured airspaces serve as a vital means of monitoring the airspace. \acrlong{atc} can effectively track \acrshort{uav}s within the corridors, ensuring compliance with regulations and identifying potential safety concerns, such as instances of proximity with other aircraft.\par
\begin{figure}[htb!]
    \centering
    \includegraphics[width=0.35\textwidth]{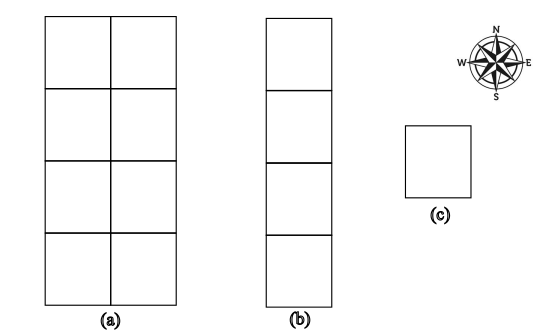}
    \caption{(a) Air corridors are composed of (b) air lanes which in turn contain (c) air cells}
    \label{fig:air corridors}
\end{figure}

Structured airspaces can be designed through the establishment of air corridors, which serve as well-defined three-dimensional pathways for enabling safe navigation for \acrshort{uas}s \cite{namuduri2022advanced}. As depicted in \Cref{fig:air corridors} air corridors are composed of air lanes \& air cells and can form complex structures such as intersections.
\subsection{Air Cells}
Air cells represent the smallest unit of discretization within the airspace and are rectangular prisms with dimensions determined by the specific characteristics of the \acrshort{uas}s being operated. These air cells facilitate the fine-grained partitioning of the airspace, enabling precise control and management of air traffic. By discretizing the airspace into these air cells, the surrounding environment becomes structured, allowing for efficient navigation and avoidance of potential conflicts. This paper assumes that each air cell measures \qty{200}{\meter} x \qty{100}{\meter} x \qty{50}{\meter} (LxBxH). 
\subsection{Air Lanes}
\begin{figure}[htb!]
    \centering
    \includegraphics[width=0.2\textwidth]{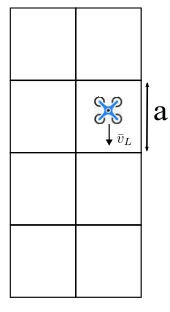}
    \caption{Air Lane}
    \label{fig:Air_Lane}
\end{figure}
Air lanes are linear collections of air cells that form unidirectional pathways within the airspace. These lanes enable the controlled movement of \acrshort{uav}s, limiting their traffic flow to a single direction. By confining \acrshort{uav}s to designated air lanes, the risk of collisions and congestion is minimized, ensuring the safe and efficient flow of air traffic. Each air lane is composed of a series of air cells, and \acrshort{uav}s navigate through these lanes to reach their intended destinations.

\subsection{Air Corridors}
Air corridors refer to collections of multiple air lanes that have \acrshort{uas}s moving parallel or anti-parallel in reference to each other. Air corridors allow for the coordinated movement of \acrshort{uav}s in a specific area along a specific line. Increasing the number of air lanes in an air corridor will increase the volume of traffic flowing along a given direction in the area. These corridors can accommodate different types of air traffic, facilitating efficient and organized \acrshort{uav} operations. Air corridors are particularly useful for managing \acrshort{uav} networks in complex environments with diverse traffic flows.

\subsection{Intersections}
\begin{figure}[htb!]
    \centering
    \includegraphics[width=0.4\textwidth]{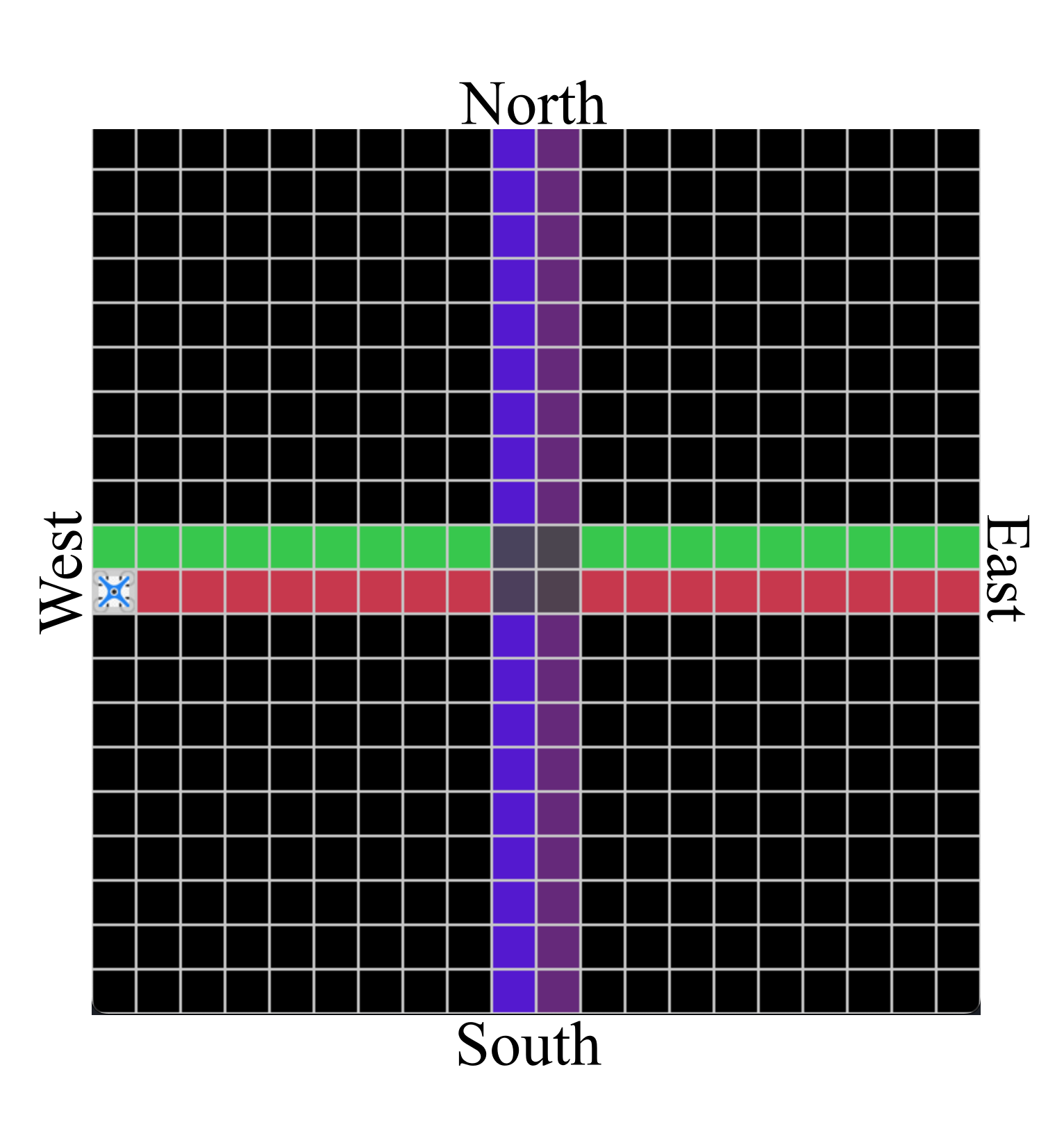}
    \caption{Green lane is Eastbound, Red  lane is Westbound, Purple lane is Northbound  \& Violet lane is Southbound. Black grids constitute a non-traversal area. Intersections are shown in grey color.}
    \label{fig:intersection}
\end{figure}
Intersections form when two or more air lanes  different directions overlap. Intersections are an inevitable feature of any transportation network that hosts a large number of vehicles. At these intersections, special protocols and coordination mechanisms are required to manage the flow of \acrshort{uav}s to prevent conflicts.
The intersection of the 2 air corridors result in a ”directionless” space that does not impose any restrictions on which directions the \acrshort{uav} can travel, as indicated in grey in the simulation visualization. This also gives \acrshort{uav}s a chance to change their directions and enter a different air lane, which may also be in a different air corridor. This intersection is analogous to a crossroad seen on the ground. 
\subsection{Vertiports}
Vertiports are takeoff and landing facilities for all \acrshort{uas} \cite{NASA-Vertiport}. All flight plans  begin from  and terminate at vertiports. Vertiports mark the entrances and exits to air corridors. 
\par
By employing a combination of air cells, air lanes, air corridors, and intersections,  structured airspaces provide a clear framework for organizing and managing \acrshort{uav} traffic, enhanced safety, seamless traffic, efficient  in \acrshort{uav} operations, and optimal \& allocation of resources.

\section{Traffic Control Framework}
\label{sec:TCF}
The \acrshort{tcf} for \acrshort{aam}  is designed utilizing the concept of structured airspaces. \acrshort{tcf} facilitates autonomous traffic management amongst independent actors. This framework helps visualize high volumes of air traffic in real time. Additionally, \acrshort{tcf} can simulate variety of air traffic conditions and estimate traffic parameters. This is achieved by enforcing a set of traffic rules on all the \acrshort{uas} that operate under this framework; providing a unified method fortraffic management. \acrshort{tcf} modularizes \acrshort{uav} behavior through the implementation of \acrshort{uas} control agents providing a high-level control that enables instant modification of the vehicles' task. The \acrlong{ta} feature of the \acrshort{tcf} details essential metrics of the traffic. An operator can then use these statistics to control the traffic further. \acrshort{ta} establishes a feedback mechanism in this framework. 

\subsection{Conformance Guidelines}
\label{sec:uav pre-req}
Every \acrshort{uav} must conform to the following guidelines which ease large-scale air traffic management. 

\begin{enumerate}[label=\roman*)]
    \item \textbf{Unique Identification:} \acrshort{faa} also requires each \acrshort{uav} to have a remote identification \cite{RemoteID}. An equivalent and simplified version of this is facilitated through an \acrfull{uid} given to each \acrshort{uav}.
    \item \textbf{Broadcasting Information:} All \acrshort{uav}s are obligated to broadcast certain environmental information and a unique identifier at predetermined intervals. This rule aligns with existing regulations set by the \acrshort{faa} for \acrshort{uav}s flying in most airspaces. This data must also be captured by each \acrshort{uas} to gain insights into other dynamic systems present in their environment.
   \item \textbf{Single Occupancy of Air Cells:} No air cell is allowed to accommodate more than one  \acrshort{uas} simultaneously. This rule ensures that there is no congestion or conflict within each discrete unit of airspace.
   \item \textbf{Single Occupancy of Intersection:} At most one \acrshort{uav} is allowed to use more than an intersection at any given point in time. This rule reduces the chances of collisions.
    \item \textbf{Safety Gap between UAVs:} It is required that each \acrshort{uav} maintains a minimum gap of at least 1 air cell when traveling within the same air lane. This rule guarantees that each vehicle has sufficient space to stop and recalibrate in the event of a malfunction or stoppage of the \acrshort{uav} ahead.
\end{enumerate}
 
\subsection{\acrshort{uas} Control Agents}
\acrshort{uas} control agents define the behaviour of the \acrshort{uav}. Technically, they are the programs that process information sensed by a \acrshort{uas} to devise subsequent actions. These \acrshort {uas} control agents allow for the extension of these abstractions, enabling support for a wide range of \acrshort{uav} controllers, thus unifying \acrshort{uas} management. Control agents receive the current state of the \acrshort{uas} and the sensed environmental data as inputs. The agents analyze it in order to formulate subsequent actions. The method employed in this paper involves a rules-based approach; a predetermined set of rules or procedures dictate the agent's behavior. The agent assesses the current state of the \acrshort{uas} alongside its operating environment and determines the most suitable course of action as guided by its programming rules. \Cref{tab:action to movement} shows the rules implemented. In addition to the rules shown, a Not-Operating (NOP) state is also defined in which no operation is performed. This system of discrete actions simplifies the decision-making process while maintaining sufficient flexibility for effective \acrshort{uav} navigation within the discrete and fully observable environment, and breaks down complex \acrshort{uas} operations into manageable and understandable units. 

\begin{table}[H]
    \centering
    \begin{tabular}{|c|c|c|}
        \hline
       Action: RIGHT  & LEFT  & REVERSE\\  
       \hline
        N $\rightarrow$ E &  N $\rightarrow$ W  &  N $\rightarrow$ S \\
        \hline
        E $\rightarrow$ S &  E $\rightarrow$ N  &  E $\rightarrow$ W \\
        \hline
        S $\rightarrow$ W  & S $\rightarrow$ E  &  S $\rightarrow$ N \\
        \hline
        W $\rightarrow$ N  & W $\rightarrow$ S  &  W $\rightarrow$ E \\
        \hline
    \end{tabular}
    \vspace{1.5mm}
    \caption{Action to movement conversion table.\\N $=$ NORTH, S $=$ SOUTH, E $=$ EAST \& W $=$ WEST }
    \label{tab:action to movement}
\end{table}

\subsection{Traffic Performance Indicators}
This paper is focused on building a traffic management model and therefore it is important to identify some traffic performance indicators that quantize the effectiveness and usefulness of a traffic management model. \acrshort{tpi}s can also be used to compare different models of traffic generation, UAS Control Agents as well as different environment parameters. \acrshort{tpi} facilitate a feedback mechanism into the Framework; whereby operators receive critical information about the traffic behavior required to manage it. A comprehensive list of \acrshort{tpi} is presented below. 

\begin{enumerate}
    \item \textbf{Mean UAV Velocity ($v_{\text{mean}}$):} Mean UAV velocity is the velocity that each UAV is expected to attain in smooth traffic flow. Given  $N_{uav}$ is the number of UAVs and $v_i$ is the velocity of the $i^{th}$ UAV, then the mean velocity \cite{Bellomo2002traffic} is defined as, $$v_{\text{mean}}= \frac{1}{N_{uav}} \sum\limits_{i\in N_{uav}} v_i.$$
    
    \item \textbf{Traffic Smoothness ($\sigma_v$):} Traffic smoothness is the measure of the small-scale variations in traffic flow. Numerically, it can be defined as the standard deviation of UAV velocity vector within the observation space and is defined as,  $$\sigma_v=
    \sqrt{\frac{1}{N_{uav}}\sum\limits_{i\in N_{uav}} \left(v_i - v_{\text{mean}}\right)^2.}$$ 

     \item \textbf{Traffic Delay ($T_{\text{td}}$):} Traffic delay is the amount of time an \acrshort{uav} waits until the gap in front of it is greater than the safety gap. 
     
    \item \textbf{Congestion Delay ($T_{\text{cong}}$):} Congestion is the situation when a vehicle is unable to move forward. Congestion delay is the amount of time an \acrshort{uav} waits until the gap in front of it is greater than the safety gap. 

\end{enumerate}

\section{Simulations and Results}
\label{sec:sim and result}
The \acrshort{tcf} is primarily a tool that can be used to check whether a given strategy on a \acrshort{uav} controller is effective under ideal conditions. This in turn allows us to work on the control logic of the \acrshort{uav} whilst being indifferent to the hardware. We use this framework to develop a \acrshort{uav} agent that can handle a single air corridor intersection. 
This section highlights the simulation capabilities of \acrshort{tcf} and the features therein. The simulation is performed on a single intersection of the air corridor. In order to overcome the which uses a distributed queue model for conflict resolution. Finally, \acrshort{uav} traffic simulations with two different traffic densities have been performed that demonstrate traffic congestion. 

\subsection{Traffic Generation}
The current \acrshort{tcf} model follows a binary probability distribution mechanism for traffic generation, however, more complex models will be addressed in future versions. 

At a given coordinate, dispatch vertiports generate \acrshort{uav}s in a particular direction and with a particular priority. Dispatch vertiports can have different probabilistic distributions powering them, and currently use the binary periodic distribution. The unit will attempt to generate a \acrshort{uav} 60\% of the time every 5 epochs (the probability and the time period can be changed), and if the binary periodic distribution returns true it will dispatch a \acrshort{uav} on the map.

\subsection{Congestion}
Congestion refers to the scenario where traffic flow is not continuous. In the model used in this paper, congestion is caused when \acrshort{uav}s stop in order to maintain the safety gap. As traffic generation rate is the only factor affecting congestion, the threshold rate was investigated. Additionally, the traffic characteristics were analysed using the \acrshort{tpi}s.

\begin{figure}[htb!]
    \centering
    \includegraphics[width=0.4\textwidth]{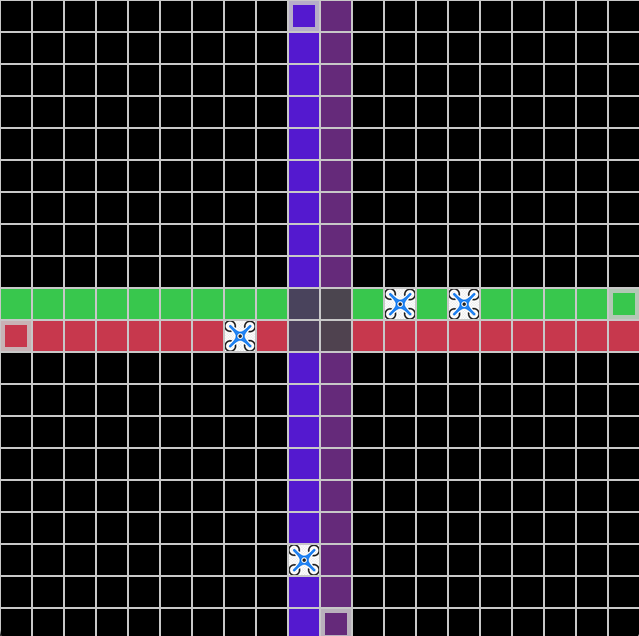}
    \includegraphics[width=0.48\textwidth]{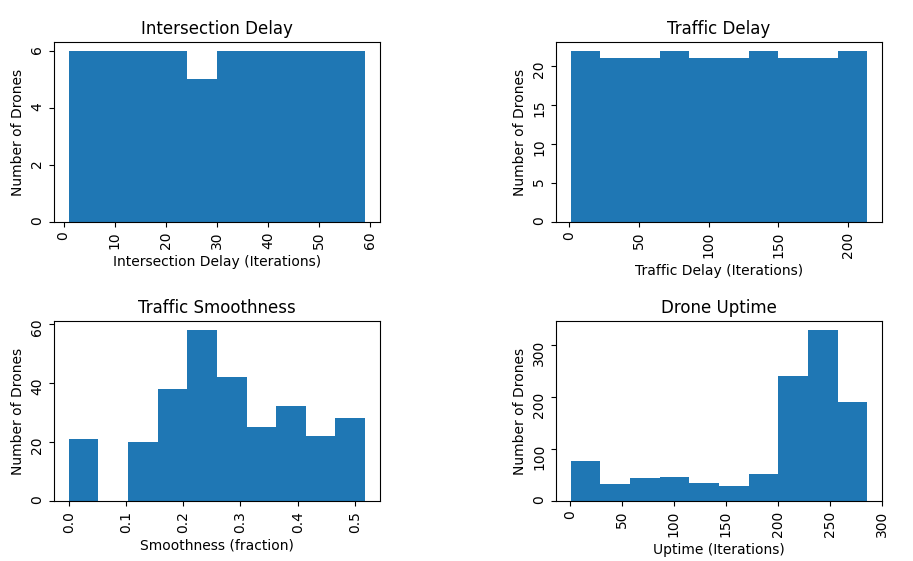}
    \caption{Non-Congested Traffic Simulation}
    \label{fig:low-non-cong-sim}
\end{figure}

It was observed that traffic generation rate of 0.25 is the threshold rate for congestion. When the traffic generation is set to less than 0.25, traffic congestion is never present. \Cref{fig:low-non-cong-sim} shows a visual representation of non-congested traffic. 


\begin{figure}[htb!]
    \centering
    \includegraphics[width=0.4\textwidth]{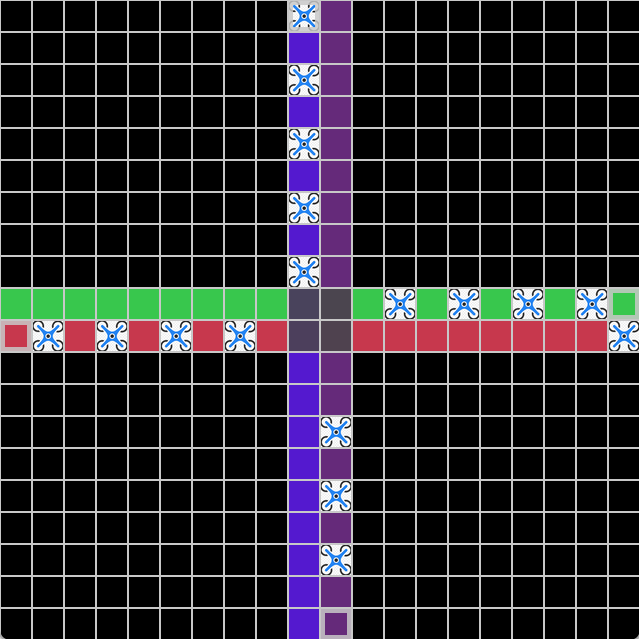}
    \includegraphics[width=0.48\textwidth]{Images/Low-Congested-Graph.png}
    \caption{Congested Traffic Simulation}
    \label{fig:low-cong-sim}
\end{figure}

When the traffic generation is set to more than 0.25, traffic congestion is never present. \Cref{fig:low-cong-sim} shows a visual representation of non-congested traffic. 

This is still a low traffic generation rate, however it is congesting since the bottleneck is 0.25 and the generation rate can be maximum up to 0.3. This will gradually cause a build up of drones in the system till there is a congestion.

\section{Conclusion}
\label{sec:conclusion}
Structured airspaces impose certain restrictions, enabling safer and more efficient deployment of a large variety of \acrshort{uav}. The idea of air corridors - with its constituent air lanes, and air cells - defines one such structure for airspaces. This concept was further enhanced by augmenting intersections and vertiports into it. Intersections are a natural consequence of structured transportation networks as any collection of \acrshort{uav}s will eventually cross each other's path. In the case of vertiports, they are the terminal points of a \acrshort{uav}'s path. Thus, these two features make the air corridors model more substantial.

However, these structures are only pragmatic when they have an accompanying set of guidelines. To this extent, a novel \acrshort{tcf} is proposed that seeks to abstract and unify the deployment, tracking, and controlling of autonomous air traffic. \acrshort{tcf} defines \acrshort{uas} control agents which computerises a real-life \acrshort{uav} along with its control logic. This digital identity can then be used for a varied set of operations whilst being agnostic to hardware. \acrshort{tpi}, described within \acrshort{tcf}, disseminates certain information about the traffic facilitating decision-making, should the need arise. The simulation feature of \acrshort{tcf} enables ahead-of-time visualization of the aerial traffic which is essential in dispelling any deployment-related concerns. Using the very same feature, two scenarios were simulated as a demonstration of its capabilities, which test the traffic congestion. 
\section{Future Scope}




To achieve more sophisticated decision-making and improve adaptability, the future version of the \acrshort{tcf} will leverage advanced AI/ML models. Techniques such as Q-Learning will enable the agents to learn optimal decision-making policies based on their experiences and feedback from the environment. Unlike the strict, deterministic nature of the current rules-based approach, this AI-driven method will allow \acrshort{uas} to evolve and refine their decision-making capabilities over time, becoming more adept at handling complex and dynamic scenarios.

\textbf{Data Availability Statement} An implementation of \acrshort{tcf} can be found on out github page \href{https://github.com/Ananay-22/UTM}{https://github.com/Ananay-22/UTM}

\bibliographystyle{unsrt}
\bibliography{references}
\end{document}